\documentclass[runningheads]{llncs}

\usepackage[T1]{fontenc}
\usepackage{algorithm}
\usepackage{algpseudocode}
\usepackage{algpseudocode}
\usepackage{amsmath}
\usepackage{amssymb}
\usepackage{mathtools}
\usepackage{amsfonts}
\usepackage{graphicx}
\usepackage{multirow}
\usepackage{makecell}
\usepackage{booktabs}
\usepackage{bbding}
\usepackage{adjustbox}
\usepackage{float}

\begin{document}
\title{ReFineG: Synergizing Small Supervised Models and LLMs for Low-Resource Grounded Multimodal NER}
\titlerunning{ReFineG for Low-Resource GMNER}

% 正式版本
\author{
Jielong Tang\inst{1,2}\thanks{These authors contributed equally.} \and
Shuang Wang\inst{3}\protect\footnotemark[1] \and
Zhenxing Wang\inst{4}\protect\footnotemark[1] \and
Jianxing Yu\inst{1,2} \and
Jian Yin\inst{1,2}\Envelope
}
\authorrunning{Tang et al.}

\institute{School of Artificial Intelligence, Sun Yat-sen University\and
Key Laboratory of Sustainable Tourism Smart Assessment Technology, Ministry of Culture and Tourism, Sun Yat-sen University \and
Beijing Normal University \and
Institute of Software, Chinese Academy of Sciences\\
\email{tangjlong3@mail2.sysu.edu.cn, issjyin@mail.sysu.edu.cn} }

% 匿名写法
% \author{Anonymous Submission}
% \authorrunning{Anonymous}

%
\maketitle              % typeset the header of the contribution
\begin{abstract}
Grounded Multimodal Named Entity Recognition (GMNER) extends traditional NER by jointly detecting textual mentions and grounding them to visual regions. While existing supervised methods achieve strong performance, they rely on costly multimodal annotations and often underperform in low-resource domains. Multimodal Large Language Models (MLLMs) show strong generalization but suffer from Domain Knowledge Conflict, producing redundant or incorrect mentions for domain-specific entities. To address these challenges, we propose t\textbf{R}aining-r\textbf{EFINE}ment-\textbf{G}rounding (\textbf{ReFineG}), a three-stage collaborative framework that integrates small supervised models with frozen MLLMs for low-resource GMNER. In the Training Stage, a domain-aware NER data synthesis strategy transfers LLM knowledge to small models with supervised training while avoiding domain knowledge conflicts. In the Refinement Stage, an uncertainty-based mechanism retains confident predictions from supervised models and delegates uncertain ones to the MLLM. In the Grounding Stage, a multimodal context selection algorithm enhances visual grounding through analogical reasoning. In the CCKS2025 GMNER Shared Task, ReFineG ranked second with an F1 score of 0.6461 on the online leaderboard, demonstrating its effectiveness with limited annotations. The code and data will be available at \url{https://github.com/tangjielong928/ReFineG}

\keywords{Multimodal Named Entity Recognition  \and MLLMs}
\end{abstract}
\section{Introduction}
Grounded Multimodal Named Entity Recognition (GMNER) is an emerging information extraction task that aims to jointly identify entity mentions of predefined types and their corresponding visual regions from text–image pairs. It has broad applications in domains such as multimodal knowledge graph construction~\cite{TIVA-KG}, visual question answering~\cite{li2025answering}, etc. However, existing multimodal NER datasets (e.g., Twitter-15~\cite{twitter15}, Twitter-17~\cite{twitter17}) lack explicit visual entity annotations, making it difficult to align textual entities with visual regions. To bridge this research gap, Yu et al.~\cite{gmner} extended these datasets with extensive fine-grained annotations of visual entities and proposed a new GMNER benchmark. Building upon this foundation, a variety of supervised approaches have since been introduced, including sequence generation~\cite{fg-gmner,gmner}, set matching~\cite{mqspn}, sequence labeling~\cite{li2024llms}, and knowledge-enhanced models~\cite{ok2024scanner}. 

\begin{figure}[h]
\vspace{-10pt}
\centering
\includegraphics[width=1.0\linewidth]{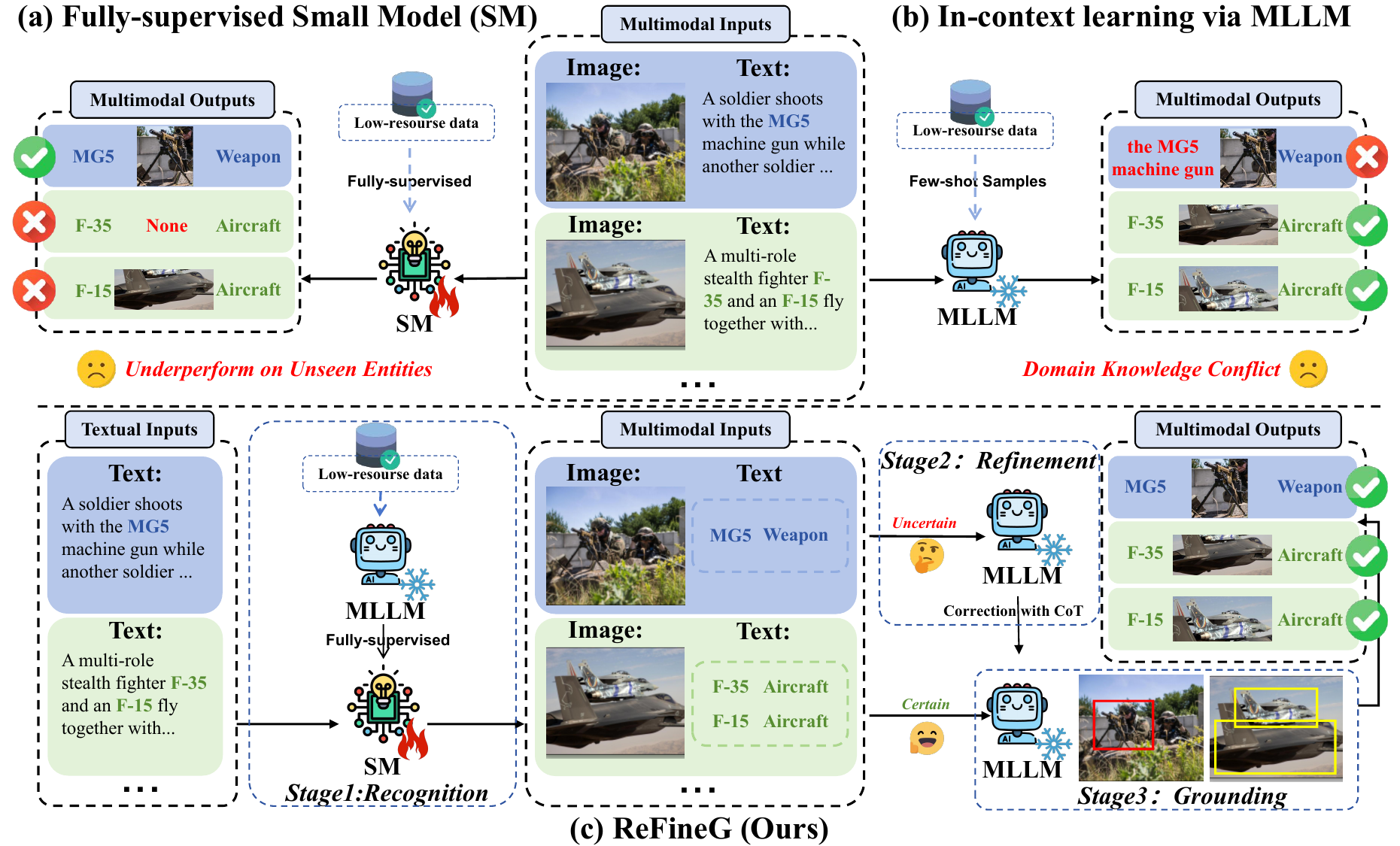}
\caption{The comparison of existing approaches and our \textbf{ReFineG}.}
\label{fig:intro}
\vspace{-15pt}
\end{figure}

Despite their effectiveness, these methods typically rely on large-scale training data, which limits their applications to real-world scenarios. In vertical domains such as military and medical, multimodal annotations, especially for visual entities, require strong domain expertise, making large-scale labeling costly and impractical. As shown in Fig.~\ref{fig:intro} (a), fully supervised methods often underfit in low-resource military domain, failing to recognize \textbf{Unseen Entities}, such as the visual regions of F-35 and F-15. With the rapid advances in multimodal pretraining and model scaling, multimodal large language models (MLLMs) have demonstrated extensive knowledge and strong visual generalization capabilities, making them promising solutions in low-resource settings. However, directly applying MLLMs to GMNER task remains challenging, as their generic knowledge often conflicts with domain-specific knowledge—a phenomenon we term \textbf{Domain Knowledge Conflict}. This issue typically arises in textual entity recognition. As illustrated in Fig.~\ref{fig:intro} (b), in a military dataset, MLLMs will produce redundant generic mentions for textual entity (e.g., “the MG5 machine gun”). Such intrinsic discrepancies in knowledge understanding are difficult to resolve through a simple prompt design~\cite{han2023information}. In contrast, these domain-specific knowledge can be effectively captured by supervised models when data is sufficient. Considering that fine-tuning an MLLM still requires substantial computational resources, we pose the question: \textit{Can small-scale supervised models be leveraged to resolve Domain Knowledge Conflict in textual entities, while MLLMs/LLMs contribute their knowledge and generalization ability to tackle unseen entities in low-resource settings?}

To this end, as shown in Fig.~\ref{fig:intro} (c), we propose a three-stage collaborative framework, t\textbf{R}aining-r\textbf{EFINE}ment-\textbf{G}rounding (\textbf{ReFineG}), where the small supervised model and the freezing MLLM complement each other’s capabilities for low-resource GMNER task. Specifically, \textbf{in the Training Stage}, we first introduce a domain knowledge-aware NER data synthesis strategy. This method maintains a global annotation guideline table, allowing the LLM to traverse the limited labeled data, summarize annotation details, and store them. Additional training data are then generated based on this guideline, which is subsequently used to train a small supervised model. Unlike directly prompting the LLM with the guideline, this approach avoids the Domain Knowledge Conflict that arises when LLMs output textual entities directly, while enabling the small model to acquire additional knowledge from the LLM. \textbf{In the Refinement Stage}, we design an uncertainty-based refinement mechanism, where the small supervised model retains high-confidence predictions and passes uncertain predictions to the LLM for further refinement. This enables effective collaboration between the two models: the supervised model handles domain-informed entities, while the LLM focuses on uncertain entities. \textbf{In the Grounding Stage}, the MLLM leverages textual entity results for visual entity grounding. To address potential Domain Knowledge Conflict, we propose a multimodal in-context sample selection algorithm, which selects the most appropriate in-context examples from entity, sentence, and image levels. This enhances the MLLM’s ability to perform visual analogical reasoning using similar examples, thereby improving the accuracy of visual entity grounding. 

Our contributions are summarized as follows: (1) We identify the challenges of low-resource GMNER task, highlighting both the Domain Knowledge Conflict in MLLMs and the underfitting of small supervised models. (2) We propose \textbf{ReFineG}, a three-stage collaborative framework that effectively integrates small supervised models with frozen MLLMs to overcome these limitations. (3) Extensive experiments demonstrate the effectiveness of our approach, ranking second place on the CCKS2025 GMNER Shared Task.

% 具体而言，In Recognition Stage, 我们首先提出了一种domain guideline-aware的NER数据生成方法，其维护了一个全局标注guideline表，允许LLM遍历有限的有标注数据总结标注细节并储存。基于此guideline进行额外训练数据生成。随后，一个序列生成模型基于生成数据被训练。不同于直接利用guideline提示LLM，这种方式避免了MLLM直接输出文本实体结果而产生的Domain Knowledge Conflict问题同时使得小模型学习到了MLLM的额外知识。In Refinement Stage，一种基于不确定性的修正机制被设计，在此小监督模型保留他们置信度高的预测结果而将他们高不确定性的传递给LLM进行进一步修正。这使得两个模型相互协同：监督模型处理已知的实体而LLM处理unseen实体。In Grounding Stage，MLLM利用文本实体结果进行视觉实体定位。由于大模型本身的知识依然可能缺失，因此我们提出了一种多模态上下文样例选择算法，其从实体，句子，图片三个角度进行最相似上下文样例的选择，提升MLLM利用相似实体进行视觉类比推理的能力，有效提高视觉实体定位性能。

\section{Related Work}
Unlike prior multimodal NER studies~\cite{twitter17,twitter15} that focus only on textual mentions, GMNER seeks to extract multimodal entity information, including entity mention, type, and corresponding visual region from text–image pairs. Existing methods, including sequence labeling~\cite{li2024llms}, sequence generation~\cite{fg-gmner,gmner}, and set-matching~\cite{mqspn}, predict mention–type–region triplets by fine-tuning Transformer-based models~\cite{bert,lewis2019bart} on large annotated GMNER datasets. While effective, they perform poorly in low-resource domains, where alignment between text and visual regions often suffers underfitting. Some works have explored knowledge enhancement, such as web retrieval~\cite{wang2022named,ok2024scanner}, LLMs~\cite{li2024llms,wang2024granular}, and multimodal data augmentation~\cite{li2024generative}. However, retrieved and augmented data usually introduces noise, while domain-specific inconsistencies in LLMs often lead to redundant or inaccurate mentions. Tang et al.~\cite{tang2025unco} attempts to address the insufficient domain knowledge of LLMs and the poor generalization ability of small models through a collaborative framework. However, it typically relies on extensively annotated data. In contrast, our ReFineG framework combines small supervised models with frozen MLLMs to address annotation scarcity and domain knowledge conflicts in GMNER.

\section{ReFineG Framework}

\textbf{Task Definition.} Given a text sequence $T$ and its paired image $I$, the GMNER task requires predicting a set of triplets $\{(e_k, c_k, r_k)\}_{k=1}^N$, where $e_k$ is the $k$-th entity mention, $c_k$ denotes its entity category, and $r_k$ corresponds to the 4D bounding box of the entity in $I$. If the entity is absent from the image, $r_k$ is set to $\texttt{None}$.

\begin{figure}[h]
\vspace{-15pt}
\centering
\includegraphics[width=1.0\linewidth]{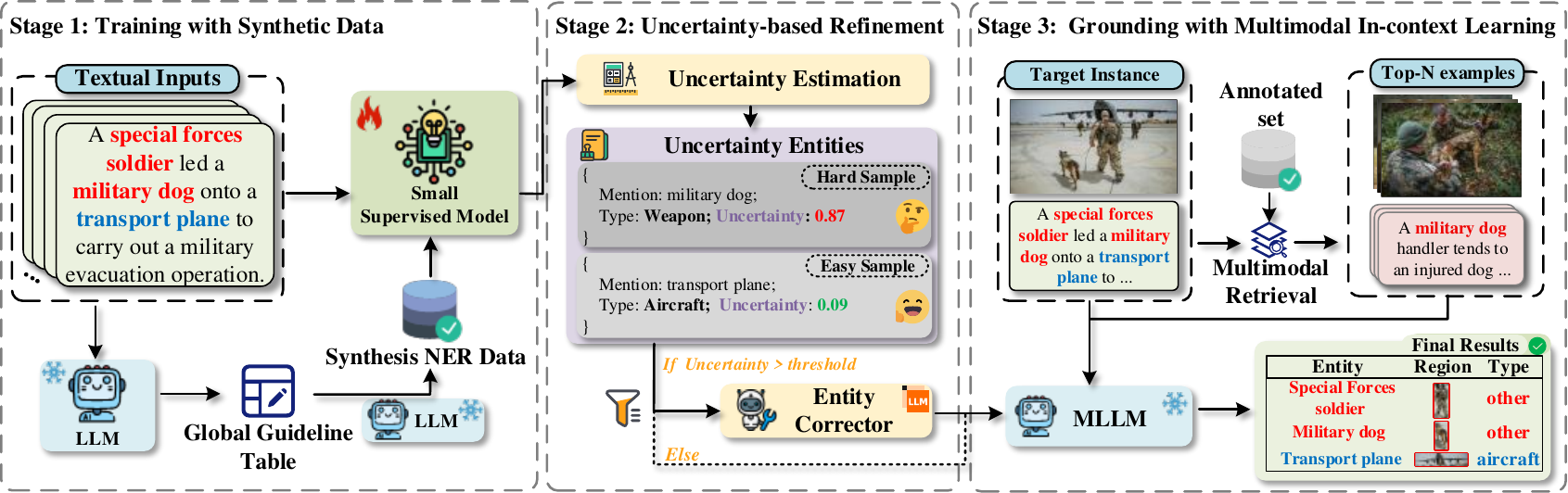}
\caption{The overall framework of our \textbf{ReFineG}.}
\label{fig:main}
\vspace{-15pt}
\end{figure}

\textbf{Overall Workflow.} As illustrated in Fig.~\ref{fig:main}, our ReFineG framework consists of three stages. (1) \textit{Training Stage:} we first introduce a domain knowledge-aware NER data synthesis strategy, which maintains a global schema guideline. The LLM traverses the limited labeled data to summarize schema details and stores them in a guideline table. Additional training samples are then generated based on this guideline, which are used to train a supervised sequence labeling model. (2) \textit{Refinement Stage:} the small supervised model is first applied to directly infer test samples. An uncertainty-based refinement mechanism allows the model to retain high-confidence predictions while passing uncertain cases to the LLM for further correction. The final textual NER output integrates results from both models. (3) \textit{Grounding Stage:} the integrated textual predictions are fed into an MLLM acting as an entity region detector. To address potential domain knowledge gaps, a multimodal in-context selection algorithm selects the most relevant top-N examples, which are then combined with the multimodal input for few-shot visual entity grounding.

\begin{figure}[h]
\centering
\includegraphics[width=1.0\linewidth]{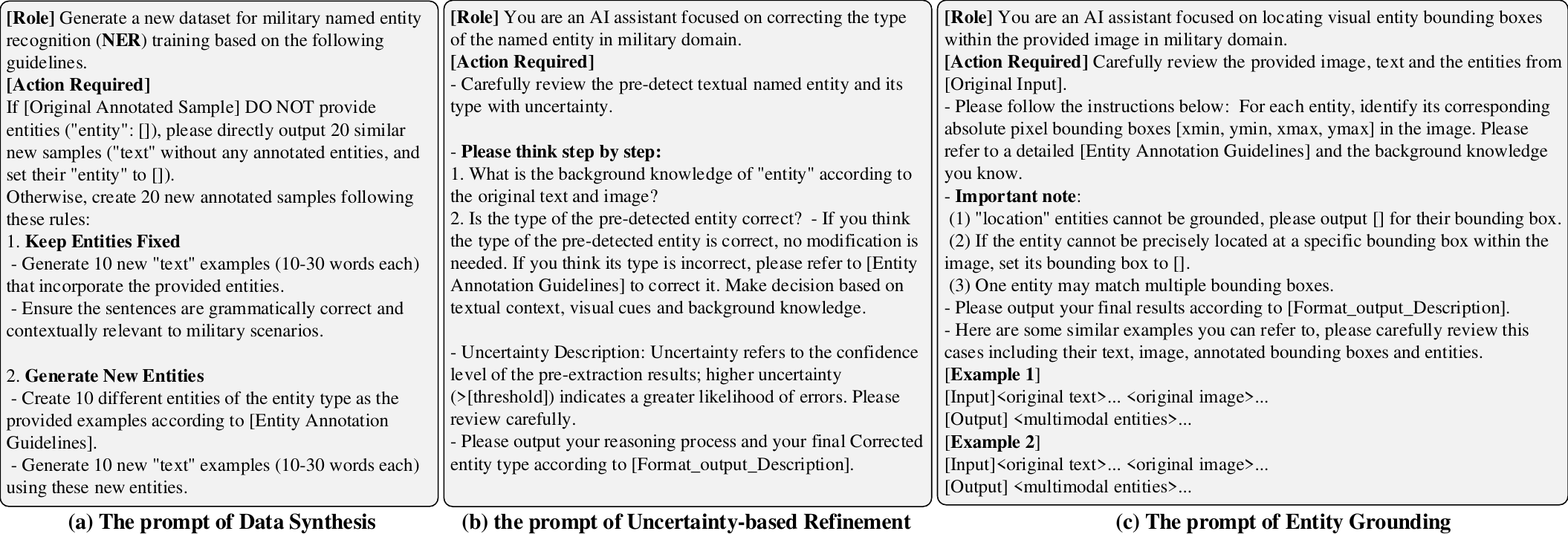}
\caption{The prompt details designed for data synthesis, refinement, and entity grounding.}
\label{fig:prompt}
\vspace{-15pt}
\end{figure}

\subsection{Stage 1: Training with Synthetic Data}
\textbf{Domain Knowledge-aware NER Data Synthesis.} 
Prior supervised approaches heavily rely on large-scale annotations and often degrade in low-resource settings. To mitigate this, we propose Domain Knowledge-aware NER Data Synthesis. Given a low-resource multimodal annotated set $\mathcal{D}=\{(s_i;e_i;v_i)\}_{i=1}^{|\mathcal{D}|}$, where $s_i$ denotes a sentence, $e_i$ is labeled entities, and $v_i$ indicate the corresponding image with annotated bounding boxes, we construct domain-specific schema guidelines. Specifically, we maintain a global guideline table $\mathcal{G}=\{Typ.; Des.; Neg.\}$, where $Typ.$ indicates the entity schema, $Des.$ provides its schema description, and $Neg.$ includes LLM-generated negative sample descriptions to guide boundary cases. $\mathcal{G}$ is dynamically updated by prompting an LLM to traverse the limited annotations in $\mathcal{D}$. We formulate it as following:
\begin{table}[h]
\vspace{-15pt}
    \fontsize{9}{9}\selectfont
    \centerline{
    \begin{tabular}{|l|}
    \hline
    \textbf{Update \{$Neg.$\}}: \texttt{Given \{$\mathcal{G}$\} and \{$s_i$\}, provide your NER results.}\\ 
    \texttt{Compare them with \{$e_i$\}, analyze error samples and update \{$Neg.$\}.}\\
    \textbf{Update \{$Des.$\}}: \texttt{Given \{$\mathcal{G}$\} and \{$(s_i;e_i)$\}, analyze the}\\ 
    \texttt{guideline of \{$e_i$\} carefully. Update \{$Des.$\} accordingly.}\\
    % \textbf{Task}: \textcolor{blue}{$\{Image\ \allowbreak Caption\}$} \\
    % \textbf{Text}: \textcolor{blue}{$\{Sentence\}$} \\
    % \textbf{Question}: \texttt{In the context of the provided information, tell me briefly what is}\\
    % \texttt{information, tell me briefly what is}\\
    % \texttt{the \textcolor{blue}{$\{Named\ \allowbreak  Entity\}$} in the Text?} \\
    % \textbf{Answer}: \textcolor{blue}{$\{Entity\ \allowbreak Expansion\ \allowbreak  Expression\}$} \\
    \hline
    \end{tabular} {}}
    \vspace{-15pt}
\end{table}
\noindent

After constructing $\mathcal{G}$, we adopt it as the guideline for prompting the LLM (as shown in Fig.~\ref{fig:prompt} (a)) and design two data synthesis strategies: (1) \textit{Entity-level Substitution}, where new entity mentions consistent with $\mathcal{G}$ are injected while preserving the original semantics and structure; (2) \textit{Sentence-level Paraphrasing}, where sentences are rephrased with contextual variations while keeping entity mentions unchanged. Finally, we obtain the synthesized dataset $\mathcal{D^{*}}=\{(s^{*}_i;e^{*}_i)\}_{i=1}^{|\mathcal{D^{*}}|}$

\textbf{Named Entity Recognition Training.} Han et al.~\cite{han2023information} have shown that in-context learning paradigm often fails to deliver satisfactory NER results, primarily because the limited input length of LLMs restricts access to sufficient domain knowledge. To mitigate domain knowledge conflicts, we further train a small supervised model on the synthesized data. We first obtain representations of the input text sequence, $\{w_1, \ldots, w_n\}$, by feeding it into a Transformer-based encoder. A subsequent linear-chain Conditional Random Field (CRF)~\cite{lafferty2001conditional} layer is then applied to predict the target label sequence $y = \{y_1, \ldots, y_n\}$:
\begin{equation}
P(y|s^{*})=\frac{\prod_{i=1}^n\psi(y_{i-1},y_i,w_i)}{\sum_{y^{\prime}\in C}\prod_{i=1}^n\psi(y_{i-1}^{\prime},y_i^{\prime},w_i)}
\end{equation}
where $\psi$ denotes the potential function, and $C$ represents the set of all possible label sequences. The negative log-likelihood (NLL) loss is used for final training:
\begin{equation}\mathcal{L}(\theta)=-\log P_\theta(\boldsymbol{e}^*|s^{*})\end{equation}

\subsection{Stage 2: Uncertainty-based Refinement}
Due to the limited generalization of the small supervised model, it inevitably produces errors, especially for entities unseen in the training set. To mitigate this, we design an uncertainty-based refinement mechanism. Given the trained small supervised model, the token-level entropy are computed as:
\begin{equation}
    \mathrm{Entropy}(w_i)=\sum_{c\in C}-P(y_i=c|T)\log P(y_i=c|T)
\end{equation}
Based on the above equation, we define the entity-level uncertainty\footnote{Maximum entropy arises when all token labels are equally likely, indicating the model’s uncertainty \cite{zhang2024linkner}.} as:
\begin{equation}
    \mathrm{Uncertainty}(e_i)=\frac{\sum_{w_i\in e_i} \mathrm{Entropy}(w_i)}{|e_i|}
\end{equation}
where $|e_i|$ denotes the length of the entity mention $e_i$. Higher uncertainty indicates lower confidence of the small supervised model in its predictions. We set a threshold $\beta$ ($\mathrm{Uncertainty}(e_i) > \beta$) to filter highly uncertain predictions, which are then passed, along with the corresponding image, to the MLLM for refinement. In this work, $\beta$ is set to 0.8. Additionally, we leverage chain-of-thought (CoT)~\cite{wei2022chain} reasoning to guide the MLLM through step-by-step corrections. The prompt design of MLLM refinement is shown in Fig.~\ref{fig:prompt} (b).

\subsection{Stage 3: Grounding with Multimodal In-context Learning}
Considering the challenges of aligning fine-grained visual regions with entities under low-resource conditions, we employ in-context learning for MLLM to perform entity grounding. In-context examples provide MLLM with direct domain knowledge, and the relevance between these examples and the target instance largely determines grounding performance~\cite{liu2021makes}. To this end, we propose a \textit{multimodal in-context examples selection algorithm} to choose the most relevant examples for in-context learning. Given the limited annotated set $\mathcal{D}=\{(s_i;e_i;v_i)\}_{i=1}^{|\mathcal{D}|}$ and the refined output $\mathcal{R}=\{(s'_j;e'_j;v'_j)\}_{j=1}^{|\mathcal{R}|}$, where $s'_j$,$e'_j$, and $v'_j$ denote the raw sentence, refined entities, and raw image, respectively, we compute similarity matrices at the entity, text, and image levels. Let $\mathcal{M}_{text}$ and $\mathcal{M}_{image}$ represent the pre-trained encoders for the textual and visual modalities, respectively.

\textbf{Entity-level Similarity Matrix.} For entities $e_i$ from annotated set and $e'_j$ from refined output, we define a type-aware entity similarity matrix as:
\begin{equation}
\mathbf{S}_{\text{entity}}(i,j) = \frac{\mathbf{H}_{E}^{\mathcal{D}} \left(\mathbf{H}_{E}^{\mathcal{R}}\right)^\top }{||\mathbf{H}_{E}^{\mathcal{D}}||_2||\mathbf{H}_{E}^{\mathcal{R}}||_2}
+ \delta \mathbf{M},\quad \mathbf{H}_{E}^{\mathcal{D}} = \mathcal{M}_{text}(e_i),\;
\mathbf{H}_{E}^{\mathcal{R}} = \mathcal{M}_{text}(e'_j).
\end{equation}
where $\mathbf{M}_{i,j}=1$ if $\mathrm{Type}(e_i)=\mathrm{Type}(e'_j)$ and $0$ otherwise.  $\delta$ is a type-consistency margin, which is set to 0.6. Entity-level similarity is obtained by aggregating over the corresponding entity mention ranges.

\textbf{Sentence-level Similarity Matrix.} For sentence $s_i$ and $s'_j$, we define a sentence similarity matrix as:
\begin{equation}
\mathbf{S}_{\text{sentence}}(i,j) = \frac{\mathbf{H}_{S}^{\mathcal{D}} \left(\mathbf{H}_{S}^{\mathcal{R}}\right)^\top }{||\mathbf{H}_{S}^{\mathcal{D}}||_2||\mathbf{H}_{S}^{\mathcal{R}}||_2},\quad \mathbf{H}_{S}^{\mathcal{D}} = \mathcal{M}_{text}(s_i),\;
\mathbf{H}_{S}^{\mathcal{R}} = \mathcal{M}_{text}(s'_j).
\end{equation}

\textbf{Image-level Similarity Matrix.} For image $v_i$ and $v'_j$, we define an image similarity matrix as:
\begin{equation}
\mathbf{S}_{\text{image}}(i,j) = \frac{\mathbf{H}_{I}^{\mathcal{D}} \left(\mathbf{H}_{I}^{\mathcal{R}}\right)^\top }{||\mathbf{H}_{I}^{\mathcal{D}}||_2||\mathbf{H}_{I}^{\mathcal{R}}||_2},\quad \mathbf{H}_{I}^{\mathcal{D}} = \mathcal{M}_{image}(v_i),\;
\mathbf{H}_{I}^{\mathcal{R}} = \mathcal{M}_{image}(v'_j).
\end{equation}
where the images without entity regions are assigned $\mathbf{S}_{\text{image}}(i,j)=0$.

\textbf{Multimodal Examples Selection.} The multi-level similarity matrices 
$\mathbf{S}_{\text{entity}}$, $\mathbf{S}_{\text{sentence}}$, and $\mathbf{S}_{\text{image}}$  are integrated to guide in-context sample selection, ensuring that retrieved examples are both semantically and visually aligned with the target instance. The overall top-$K$ similarity matrix is calculated as:
\begin{equation}
\mathcal{I}_{overall}=\underset{i\in\{1,2,...,K\}}{\operatorname*{\mathrm{argTopK}}}\{\lambda_{1}\mathbf{S}_{\text{entity}}(i,j)+\lambda_2\mathbf{S}_{\text{sentence}}(i,j)+\lambda_3\mathbf{S}_{\text{image}}(i,j)\}
\end{equation}
where $\lambda_{1}$, $\lambda_{2}$, and $\lambda_{3}$ denote the weighting hyperparameters, $\mathcal{I}_{overall}$ is the index set of top-$K$ similar examples in $\mathcal{D}$. The multimodal in-context examples $\mathcal{C}$ are defined as follows:
\begin{equation}\mathcal{C}=\{(s_j,e_j,v_j)\mid j\in\mathcal{I}_{overall}\}\end{equation}
Finally, the prompt designed for entity grounding is illustrated in Fig.~\ref{fig:prompt} (c).

\section{Experiments}
\subsection{Experiment Setting}
\textbf{Dataset.} We conduct experiments on CCKS-GMNER in the military domain and Twitter-GMNER in the social media domain. Detailed statistics of the two datasets are presented in Table~\ref{tab:dataset}. Since CCKS-GMNER provides only 500 annotated instances, we randomly split them into 100 samples for low-resource training and 400 samples for test. To ensure a low-resource setting, we follow~\cite{li2024generative} to select the same 10\% of the Twitter-GMNER data as the training data.

\begin{table}[h]
\centering
\caption{The statistics of two GMNER datasets.}
% \begin{adjustbox}{width=1.0\columnwidth}
\begin{tabular}{l|cccc}
\toprule
    \textbf{Dataset} & \textbf{Samples} & \textbf{Types} & \textbf{Entities} & \textbf{Regions} \\
\midrule
    Twitter-GMNER & 10,000 & 4 & 16,778& 8,090 \\
    CCKS-GMNER & 500 & 6 & 384 & 307 \\
\bottomrule
\end{tabular}
% \end{adjustbox}
\label{tab:dataset}
\vspace{-10pt}
\end{table}

\textbf{Evaluation.} We adopt F1 score, Recall, and Precision as evaluation metrics. A multimodal entity is regarded as correct only when the textual entity mention, type, and visual region are all correctly predicted. For visual grounding, a region is considered correct if the Intersection-over-Union (IoU) between the predicted and ground-truth regions exceeds 0.5. For entities without visual regions, the model is required to output \texttt{None}.

\textbf{Implementation Details.} In Stage 1, all experiments are conducted on a single NVIDIA RTX 3090 GPU. We employ the Qwen-Max\footnote{\url{https://qwenlm.github.io/blog/qwen2.5-max}} API as the LLM to synthesize training data, while Bert-based\footnote{\url{https://huggingface.co/google-bert/bert-base-uncased}} and XLM-Roberta-Large\footnote{\url{https://huggingface.co/FacebookAI/xlm-roberta-large}} are adopted as the small supervised models. The learning rate is set to $1\mathrm{e}{-5}$, with a CRF layer learning rate of $5\mathrm{e}{-2}$, a batch size of 32, 10 epochs, and a maximum sequence length of 128. AdamW is used as the optimizer. In Stages 2 and 3, we utilize Qwen2.5VL-72B-Instruct\footnote{\url{https://huggingface.co/Qwen/Qwen2.5-VL-72B-Instruct}} as the MLLM, deployed on four NVIDIA A100 SXM4 80GB GPUs, with 3-shot multimodal examples for in-context learning. The pretrained models for computing text and image similarity are Sentence Transformer\footnote{\url{https://huggingface.co/sentence-transformers/all-MiniLM-L6-v2}} and CLIP-ViT-L-14\footnote{\url{https://huggingface.co/openai/clip-vit-large-patch14}}, respectively. The hyperparameters $\lambda_{1}$, $\lambda_{2}$, and $\lambda_{3}$ are set to 0.6, 0.4, and 0.2, respectively.

\textbf{Baselines.} We compare our approach against three types of baselines: (1) Fully supervised models, including \textbf{H-index}~\cite{gmner} and \textbf{MQSPN}~\cite{mqspn}, which adopt pretrained YOLOv11~\cite{khanam2024yolov11} (for CCKS\footnote{Yolov11 is pretrained as regions proposal model on open-sourced military dataset: \url{https://www.kaggle.com/datasets/rawsi18/military-assets-dataset-12-classes-yolo8-format}}) or VinVL~\cite{zhang2021vinvl} (for Twitter) as the object detector and are trained in an end-to-end manner to predict entity mention–type–region triplets; (2)MLLMs, including \textbf{Qwen2.5VL-72B}~\cite{bai2025qwen2}, and \textbf{GPT-4o}~\cite{hurst2024gpt}, where multimodal entity information is obtained via 3-shot in-context learning; and (3) Multimodal data augmentation methods, including \textbf{GMDA}~\cite{li2024generative} and \textbf{MixGen}~\cite{hao2023mixgen}, which synthesize image–text pairs to enhance the training of the H-index model.

\subsection{Main Results}
\textbf{Comparison with Previous Baselines.} Table~\ref{tab:exp_main} presents the performance comparison between our proposed ReFineG and various baseline methods. On the Twitter-GMNER dataset, ReFineG (Qwen2.5VL-72B+BERT-base) outperforms the supervised model H-index by 7.3\% F1 and surpasses GPT-4o by 13.47\% F1. On the CCKS-GMNER dataset in the military domain, ReFineG achieves superior performance over all state-of-the-art (SOTA) models. This improvement can be attributed to the collaborative framework of large and small models in ReFineG, which effectively leverages their complementary strengths. Furthermore, compared with the SOTA multimodal data augmentation method GMDA, ReFineG achieves a 5.53\% F1 gain on Twitter-GMNER. These results demonstrate that ReFineG brings substantial performance improvements under low-resource settings.

\textbf{Performance on the CCKS2025 GMNER online leaderboard.} Table~\ref{tab:exp_leaderboard} presents the performance of different ReFineG variants on the final CCKS2025@GMNER online leaderboard. The results show that each stage consistently improves over baseline. Notably, Stage~1 plays a crucial role in the overall framework, where the main gain comes from higher precision. This improvement is primarily attributed to leveraging the global schema guideline to synthesize training data, enabling the small supervised model to effectively acquire domain knowledge and thereby cooperate with the MLLM to reduce domain conflicts. Furthermore, both the refinement stage and the multimodal in-context example selection yield additional improvements, demonstrating the effectiveness of our design.

\begin{table}[h]
\vspace{-10pt}
\centering
% \small
\caption{Performance comparison of different baselines on two GMNER datasets. Bold indicates the optimal result, and underlined denotes the suboptimal result. $\clubsuit$ represents that the model results on Twitter-GMNER are from~\cite{li2024generative}. Others are implemented based on our low-resource experiment setting.}
\begin{adjustbox}{width=0.8\columnwidth}
\begin{tabular}{l|lccccccc}
\toprule
\multirow{2}{*}{\textbf{Category}} & \multirow{2}{*}{\textbf{Methods}} & \multicolumn{3}{|c|}{\textbf{Twitter-GMNER}}  & \multicolumn{3}{c}{\textbf{CCKS-GMNER}} \\
\cline{3-8}
 & &  \multicolumn{1}{|c}{\textbf{Pre.}} & \textbf{Rec.} & \multicolumn{1}{c|}{\textbf{F1}} & \textbf{Pre.} & \textbf{Rec.} & \multicolumn{1}{c}{\textbf{F1}} & \\
 \midrule 
  \multirow{2}{*}{Supervised Models} & MQSPN & \multicolumn{1}{|c}{46.28} & 43.51 & \multicolumn{1}{c|}{44.85} & 52.14 & 46.72 & 49.28 \\
  & H-index$\clubsuit$ & \multicolumn{1}{|c}{47.54} & 47.39 & \multicolumn{1}{c|}{47.46} & 49.59 & 43.47 & 46.33 \\
   \midrule 
   \multirow{2}{*}{MLLMs}  & Qwen2.5VL-72B  & \multicolumn{1}{|c}{36.85} & 45.54 & \multicolumn{1}{c|}{40.74} & 58.01 & 64.20 & 60.95 \\
   & GPT4o  & \multicolumn{1}{|c}{38.01} & 45.19 & \multicolumn{1}{c|}{41.29} & 62.17& 66.67 & 64.34 \\
   \midrule 
   \multirow{2}{*}{Data Augmentation} & MixGen$\clubsuit$ (H-index)  & \multicolumn{1}{|c}{45.72} & 50.84 & \multicolumn{1}{c|}{48.15} & -& - & - \\ 
   & GMDA$\clubsuit$ (H-index)  & \multicolumn{1}{|c}{48.99} & 49.47 & \multicolumn{1}{c|}{49.23} & -& - & - \\
   \midrule 
   \multirow{2}{*}{\textbf{Ours}} & \textbf{ReFineG (Bert-base)}  & \multicolumn{1}{|c}{\underline{52.49}} & \underline{57.23} & \multicolumn{1}{c|}{\underline{54.76}} & \underline{74.57} & \underline{71.19} & \underline{72.84} \\
   & \textbf{ReFineG (Roberta-large)}  & \multicolumn{1}{|c}{\textbf{54.13}} & \textbf{60.24} & \multicolumn{1}{c|}{\textbf{57.02}} & \textbf{76.52} & \textbf{73.09} & \textbf{74.76} \\
% $\pm$0.36 $\pm$0.41 $\pm$0.39
% & GMDA$\spadesuit$  & MM2024 & \multicolumn{1}{|c}{\underline{57.09}} & \textbf{60.21} & \multicolumn{1}{c|}{\underline{58.61}} & - & {-}& {-} & \multicolumn{1}{|c}{-} & {-} & - \\
  
\bottomrule
\end{tabular}
\end{adjustbox}
\label{tab:exp_main}
\vspace{-20pt}
\end{table}

\begin{table}[h]
\vspace{-10pt}
\centering
\small
\caption{Results of different ReFineG variants on the CCKS2025@GMNER Online Leaderboard. w/o denotes the removal of the corresponding module.}
% \begin{adjustbox}{width=0.6\columnwidth}
\begin{tabular}{l|ccc}
\toprule
\multirow{2}{*}{\textbf{Methods}} & \multicolumn{3}{c}{\textbf{Online Test}} \\
\cline{2-4}
 & \textbf{Pre.} & \textbf{Rec.} & \textbf{F1} \\
 \midrule 
  Baseline (Qwen2.5VL-72B) & 49.45 & 60.93 & 54.59 \\
  \midrule 
  ReFineG (w/o Stage 1) & 56.51 & 60.37 & 58.38 \\
  ReFineG (w/o Stage 2) & 60.20 & 63.61 & 61.86 \\
  ReFineG (w/o Stage 3) & 60.74 & 64.09 & 62.37 \\
  \textbf{ReFineG (Overall)} & \textbf{62.61} & \textbf{66.73} & \textbf{64.61} \\
\bottomrule
\end{tabular}
% \end{adjustbox}
\label{tab:exp_leaderboard}
\vspace{-10pt}
\end{table}

\subsection{Ablation Study.} 
\begin{figure}[h]
\vspace{-5pt}
\centering
\includegraphics[width=1.0\linewidth]{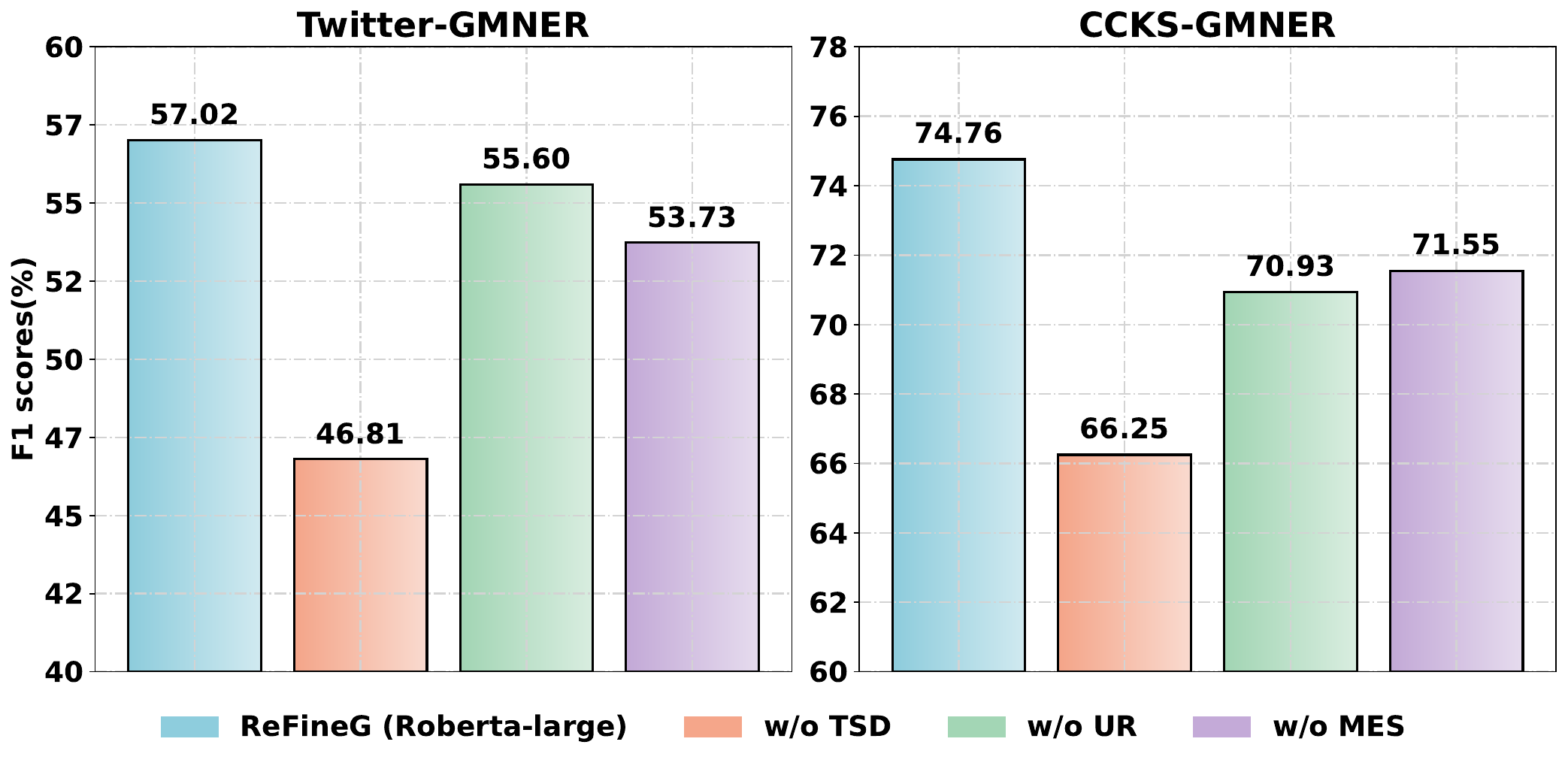}
\caption{Ablation study results. (a) Training with Synthetic Data (TSD) (b) Uncertainty-based Refinement (UR) (c) Multimodal Examples Selection (MES).  }
\label{fig:ablation}
\vspace{-15pt}
\end{figure}

To further investigate the effectiveness of each module, we conduct a series of ablation studies on the two GMNER datasets, with results shown in Fig.~\ref{fig:ablation}. (1) \textbf{w/o TSD} removes the supervised training of the synthesized dataset. We observe a substantial F1 drop on both datasets, with a decrease of 10.21\% on Twitter-GMNER and 8.51\% on CCKS-GMNER. This indicates that supervised training of the small model is essential for learning domain knowledge, while the synthetic data generated by the large model transfers its general knowledge to the small model. (2) \textbf{w/o UR} removes the MLLM refinement stage, leading to consistent performance degradation across both datasets. This demonstrates that refinement from MLLM is crucial for handling uncertain or unseen entities. (3) \textbf{w/o MES} replaces dynamic multimodal example selection with fixed 3-shot examples. The observed decline confirms that multimodal examples provide visual cues for entity grounding, enabling the MLLM to leverage analogical reasoning to improve target entity grounding.

\section*{Conclusion}
In this work, we presented ReFineG, a three-stage collaborative framework that integrates small supervised models with frozen MLLMs for low-resource grounded multimodal NER. By synthesizing domain knowledge-aware training data, leveraging uncertainty-based refinement, and performing multimodal in-context grounding, our method effectively addresses the challenges of annotation scarcity and domain knowledge conflict. Extensive experiments on Twitter-GMNER and CCKS-GMNER demonstrate that ReFineG achieves substantial improvements over state-of-the-art baselines, and ranks second on the CCKS2025 GMNER shared task leaderboard. Future directions include extending our framework to other multimodal information extraction tasks and exploring adaptive strategies for dynamically balancing the contributions of small models and MLLMs across domains.

\section*{Acknowledgements}
This work is supported by the Research Foundation of Science and Technology Plan Project of Guangzhou City (2023B01J0001, 2024B01W0004).

\bibliographystyle{splncs04}
\bibliography{custom}




% LNCS Homepage, \url{http://www.springer.com/lncs}, last accessed 2023/10/25
% \end{thebibliography}
\end{document}